\documentclass[aps,pra,amsmath,amssymb,superscriptaddress,longbibliography]{revtex4-1}
\usepackage{graphicx}
\usepackage{braket}
\usepackage{color}
\usepackage{hyperref}
\usepackage{soul}
\usepackage[dvipsnames]{xcolor}

\newcommand{\VH}{V_\mathrm{H}}
\usepackage[T1]{fontenc}

\begin{document}

\title{Gate reflectometry in dense quantum dot arrays}

\author{Fabio~Ansaloni\textsuperscript{\dag}}
\affiliation{Center for Quantum Devices, Niels Bohr Institute, University of Copenhagen, 2100 Copenhagen, Denmark}

\author{Heorhii~Bohuslavskyi\textsuperscript{\dag}}
\affiliation{Center for Quantum Devices, Niels Bohr Institute, University of Copenhagen, 2100 Copenhagen, Denmark}

\author{Federico~Fedele}
\affiliation{Center for Quantum Devices, Niels Bohr Institute, University of Copenhagen, 2100 Copenhagen, Denmark}

\author{Torbj\o rn~Rasmussen}
\affiliation{Center for Quantum Devices, Niels Bohr Institute, University of Copenhagen, 2100 Copenhagen, Denmark}

\author{Bertram~Brovang}
\affiliation{Center for Quantum Devices, Niels Bohr Institute, University of Copenhagen, 2100 Copenhagen, Denmark}

\author{Fabrizio~Berritta}
\affiliation{Center for Quantum Devices, Niels Bohr Institute, University of Copenhagen, 2100 Copenhagen, Denmark}

\author{Amber~Heskes}
\affiliation{Center for Quantum Devices, Niels Bohr Institute, University of Copenhagen, 2100 Copenhagen, Denmark}

\author{Jing~Li}
\affiliation{Universit\'e Grenoble Alpes, CEA, IRIG, MEM-L Sim, F-38000 Grenoble, France}

\author{Louis~Hutin}
\affiliation{Universit\'e Grenoble Alpes, CEA, LETI, MINATEC Campus, F-38000 Grenoble, France}

\author{Benjamin~Venitucci}
\affiliation{Universit\'e Grenoble Alpes, CEA, IRIG, MEM-L Sim, F-38000 Grenoble, France}

\author{Benoit~Bertrand}
\affiliation{Universit\'e Grenoble Alpes, CEA, LETI, MINATEC Campus, F-38000 Grenoble, France}

\author{Maud~Vinet}
\affiliation{Universit\'e Grenoble Alpes, CEA, LETI, MINATEC Campus, F-38000 Grenoble, France}

\author{Yann-Michel~Niquet}
\affiliation{Universit\'e Grenoble Alpes, CEA, IRIG, MEM-L Sim, F-38000 Grenoble, France}

\author{Anasua~Chatterjee}
\affiliation{Center for Quantum Devices, Niels Bohr Institute, University of Copenhagen, 2100 Copenhagen, Denmark}

\author{Ferdinand~Kuemmeth}
\thanks{kuemmeth@nbi.dk\\\textsuperscript{\dag}These authors contributed equally to this work.}
\affiliation{Center for Quantum Devices, Niels Bohr Institute, University of Copenhagen, 2100 Copenhagen, Denmark}

\date{\today}

\begin{abstract}
 
\textbf{Abstract:} Silicon quantum devices are maturing from academic single- and two-qubit devices to industrially-fabricated dense quantum-dot arrays, increasing operational complexity and the need for better pulsed-gate and readout techniques. We perform gate-voltage pulsing and gate-based reflectometry measurements on a dense 2$\times$2 array of silicon quantum dots fabricated in a 300-mm-wafer foundry.  
Utilizing the strong capacitive couplings within the array, it is sufficient to monitor only one gate electrode via high-frequency reflectometry to establish single-electron occupation in each of the four dots and to detect single-electron movements with high bandwidth.
A global top-gate electrode adjusts the overall tunneling times, while linear combinations of side-gate voltages yield detailed charge stability diagrams. 
To test for spin physics and Pauli spin blockade at finite magnetic fields, we implement symmetric gate-voltage pulses that directly reveal bidirectional interdot charge relaxation as a function of the detuning between two dots. 
Charge sensing within the array can be established without the involvement of adjacent electron reservoirs, important for scaling such split-gate devices towards longer 2$\times$N arrays.
Our techniques may find use in the scaling of few-dot spin-qubit devices to large-scale quantum processors. 
\end{abstract}

\maketitle

\section{Introduction} 
\label{intro}

Spin-based quantum computing based on gate-controlled silicon quantum dots is rapidly evolving~\cite{Chatterjee2021}, underscored by recent devices with single- and two-qubit fidelities exceeding 99\%~\cite{Madzik2021, Xue2021b, Noiri2021b}, all fabricated in non-industrial cleanrooms. Devices fabricated by industrial 300-mm-wafer processes already show coherent single-spin rotations~\cite{Maurand2016, Crippa2019, Li2021b, Zwerver2021}, with further efforts spanning planar Si/SiGe heterostructures~\cite{Pillarisetty2018}, planar Si/SiO$_2$ interfaces~\cite{Li2021b, Stuyck2021b}, fully-depleted silicon-on-insulator (FDSOI) silicon nanowires~\cite{Franceschi2016b, Vinet2018c} as well as $^{28}$Si/SiO$_2$ fins~\cite{Zwerver2021}. While these approaches offer the prospects of high yield and high device uniformity, device geometries are more restricted relative to academic devices, for example due to foundry preferences for photolithography over electron-beam lithography, etching over lift-off processes, and other considerations. 

Scaling spin qubits from few-qubit circuits towards fault-tolerant quantum processors will likely involve two-dimensional arrays of singly-occupied quantum dots~\cite{Loss1998, Fowler2012}, sufficiently dense to allow two-qubit gates based on Heisenberg spin exchange. 
Two-dimensional arrays have been investigated in gallium arsenide~\cite{Thalineau2012, Mortemousque2020, Mukhopadhyay2018, Fedele2021} and germanium~\cite{Riggelen2021}, all fabricated by electron-beam lithography in academic cleanrooms with dedicated proximal charge sensors that enabled the operation of these dot arrays in their one-electron regimes. A recent proposal suggests the use of sparse spin qubit arrays, in which pairs of electrons are controlled close to each other only when required~\cite{Boter2021}, although experimental advances are needed to implement and control such devices. 
 
In this work, we focus on controlling charge transitions in a foundry-fabricated 2$\times$2 silicon quadruple dot without the need for additional charge sensors. Using gate-based high-frequency reflectometry, we demonstrate charge readout, dispersive sensing, and single-electron occupation of each of the four quantum dots. We then acquire charge stability diagrams with and without compensating for capacitive crosstalk within the array, generalizing negatively compensated control voltages introduced in Ref.~\cite{Ansaloni2020} to positively compensated control voltages. This allows the acquisition of charge stability maps over wide gate-voltage regions of the qubit array, which may be useful for exploring spin qubit operations in multi-electron configurations~\cite{Leon2020}. We also extend pulsed-gate experiments from Ref.~\cite{Ansaloni2020} by  designing symmetric gate-voltage pulses that allow the detection of forward and reverse interdot charge relaxation processes, as a function of dot detuning. (In finite magnetic fields, we find no evidence for Pauli spin blockade, although similar devices have recently enabled spin-relaxation experiments~\cite{Lundberg2020, CirianoTejel2021,Oakes2022}). Finally, we show that charge sensing within the array is possible without involvement of the source or drain reservoirs. To our knowledge, this has not been reported in literature, and may inspire dense arrays of spin qubits without the need to route ohmic channels across the quantum processor. Our measurements are supported by a constant interaction model that captures multi-dot Coulomb blockade and a $\mathbf{k}\cdot\mathbf{p}$ model that confirms an overall dependence of tunnel barriers on a top-gate voltage. 
Overall, our techniques utilize the strong capacitive couplings within the dense qubit array and may be useful for scaling current spin-qubit devices to larger quantum-dot arrays.

Section~\ref{device} introduces the device and pulsed-gate reflectometry setup. 
Section~\ref{singleelectron} explains how reflectometry off one gate electrode allows detection of the first electron for all four quantum dots, albeit not simultaneously.
Section~\ref{compensated} introduces compensated control voltages for acquiring multi-dot charge stability diagrams via radio-frequency reflectometry. 
Section~\ref{pulsegate} presents the time-domain pulsed-gate measurements revealing forward and reverse charge relaxation processes across an interdot transition.
Section~\ref{molecule} implements a hybridized double dot within the array, such that nearby charge transitions can be sensed dispersively without the need for exchanging electrons with a reservoir.

\section{Methods}		
\label{device}

\begin{figure} [t!]
\includegraphics[scale=0.98]{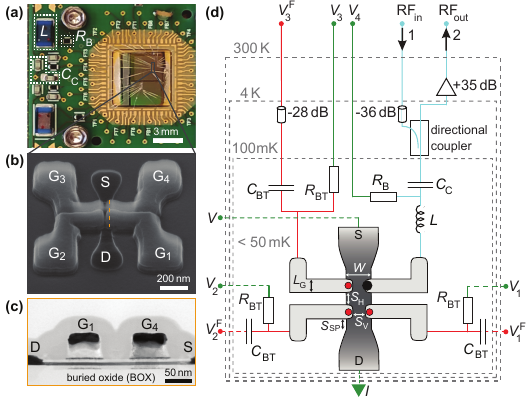}
\caption{\textbf{Device and reflectometry setup.} 
(a) Device chip wirebonded to a high-bandwidth sample holder \cite{Qdevil}. 
(b) Tilted scanning-electron micrograph of a similar quadruple dot after gate patterning. 
Accumulation gate electrodes G$_\mathrm{1-4}$ partially cover an undoped silicon nanowire between source (S) and drain (D) contacts. 
(c) Transmission-electron cross section of a fully-processed device along the nanowire as indicated in b. 
(d) Radio-frequency reflectometry setup with pulsed-gate capabilities.
Bias resistor $R_\mathrm{B}$, coupling capacitor $C_\mathrm{C}$, and inductor $L$ allow application of a tuning voltage ($V_4$) and reflectometry carrier (RF$_\mathrm{in}$) to the gate electrode of a sensor dot (G$_\mathrm{4}$). 
Bias tees ($R_\mathrm{BT}$, $C_\mathrm{BT}$) allow application of slow tuning voltages ($V_\mathrm{1,2,3}$) and fast gate-voltage pulses ($V_\mathrm{1,2,3}^\mathrm{F}$) to side gates G$_\mathrm{1,2,3}$.
The transition frequency of the bias tees is approximately 4~kHz.
	}
	\label{fig1}
\end{figure}
	 
Figure~\ref{fig1}a shows the device chip wirebonded to a high-frequency printed-circuit-board chip carrier that also provides reflectometry functionalities~\cite{Vigneau2022} via a surface-mounted inductor $L$, a coupling capacitor $C_\mathrm{C}$, and a bias resistor R$_\mathrm{B}$. 
An undoped silicon nanowire of width $W$=70~nm and thickness $t_\mathrm{Si}$=7~nm is connected to highly-doped source and drain contacts (Fig.~\ref{fig1}b). 
Four accumulation gate electrodes G$_\mathrm{1-4}$ induce electrostatically-defined quantum dots (QDs) under each gate for sufficiently positive gate voltages~\cite{Ansaloni2020}. 
Gate lengths $L_\mathrm{G}$=32~nm and vertical and horizontal spacings $S_\mathrm{V}$=$S_\mathrm{H}$=32~nm are defined by hybrid deep-ultraviolet and electron-beam lithography and silicon-nitride spacers \cite{Barraud2016,Hutin2016b}.
The gate stack comprises a 6-nm SiO$_2$ gate dielectric capped by 5-nm TiN and 50-nm doped polycrystalline silicon, as shown in Fig.~\ref{fig1}c for a similar device. 
	 
Figure~\ref{fig1}d summarizes the wiring between room temperature electronics and the device located inside a cryofree dilution refrigerator. 
The sample holder at 0.05~K allows measurements of source-drain current ($I$, measured via thermalizing filters~\cite{Qdevil}) and radio-frequency measurements via port 1 (RF$_\mathrm{in}$) and port 2 (RF$_\mathrm{out}$), building on earlier gate-based reflectometry~\cite{Volk2019a}. 

A SG383 vector source generates a reflectometry carrier that is attenuated and phase-shifted at room temperature and further attenuated inside the cryostat (36~dB distributed between 50, 4, and 1~K) before passing a directional coupler (MC ZFDC-20-5-S). 
The signal reflected from the $LC$ resonator is amplified at 4~K (35~dB, Weinreb CITLF1) and further processed at room temperature for demodulation, amplification, low-pass filtering (SR560) and digitization (AlazarTech ATS9360). Here, $L$=820~nH is a surface-mounted copper inductor (Coilcraft 1206CS-821XJL) and $C$ constitutes a parasitic capacitance to ground associated with bond wires and metal tracks (not shown in Fig.~\ref{fig1}d, estimated 0.85~pF from the observed resonance at 191~MHz). 
The 191-MHz carrier (-75~dBm incident on $L$) results in a demodulated background amplitude of 0.2~V. By adjusting the phase shifter, the demodulated quadrature $V_\mathrm{H}$ is reduced to approximately 0 and then changes by typically 0.01~V when the sensor dot is active, corresponding to a change in the phase of the reflected carrier of approximately 2$^{\circ}$.  

Side gates G$_\mathrm{1,2,3}$ are wirebonded to bias-tees so that low-frequency tuning voltages and high-frequency control pulses can be applied simultaneously. 
The high-bandwidth sample holder is available for multi-channel quantum electronics experiments~\cite{Qdevil} and features high-frequency grounds (SMD capacitors to ground) on all low-frequency channels, including source and drain wires.
For fast voltage pulses $V_\mathrm{1,2,3}^\mathrm{F}(t)$, coaxial cables are attenuated by 28~dB distributed between 50, 4, 1 and 0.05~K. 
Low-frequency (high-frequency) control voltages are generated by a high-resolution QDAC~\cite{Qdevil} (Tektronix AWG5014C), and no external magnetic field is applied unless otherwise specified.

Despite the small number of gate electrodes, tunnel rates can be adjusted in situ via a 200-nm-long copper electrode (top gate) that runs across the device center 300~nm above the nanowire (see Appendix A). 
Its wiring is identical to that of the low-pass-filtered source and drain wiring (not shown in Fig.~\ref{fig1}d). 
Alternative barrier tuning via a highly-doped silicon layer below the buried oxide (BOX) was demonstrated in Ref.~\onlinecite{Roche2012} for similar nanowire devices. 

The split-gate device studied here was recently used to demonstrate various single-electron operations at $B=0$, including the tuning of tunneling rates by the top-gate voltage ($V_\mathrm{tg}$) and the spatial permutation of two electrons~\cite{Ansaloni2020}, as well as the implementation of triggered acquisition and autonomous measurement of Coulomb blockade boundaries~\cite{Chatterjee2021a}. Similar split-gate arrays from the same foundry demonstrated capacitive coupling between two 2$\times$2 arrays~\cite{Duan2020,Gilbert2020}, microwave spectroscopy of double-dot states~\cite{Ezzouch2021}, as well as a large dispersive coupling to a microwave resonator~\cite{Ibberson2021}. In all cases, including our results below, the relatively large capacitive couplings arising from the FDSOI nanowire geometry plays an advantageous role. 

\section{Results and discussion} 
\subsection{Single-electron occupations} 
\label{singleelectron}

\begin{figure}[t]
\includegraphics[scale=0.99]{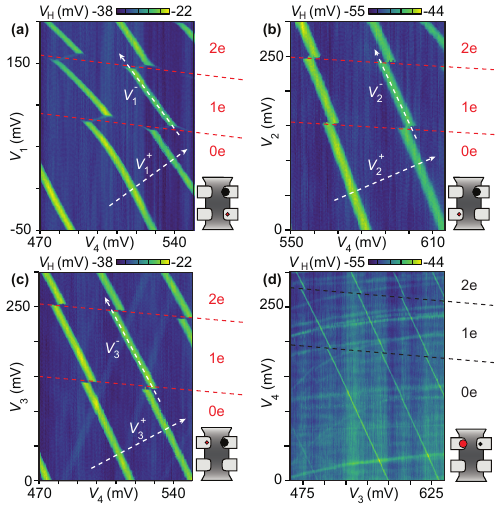}
\caption{\textbf{Single-electron counting by reflectometry}. 
Stability diagrams of QD$_{1-4}$ reveal the voltage thresholds for the first and second electron in each dot (dashed lines). 
(a-c) Peaks in the demodulated quadrature $\VH$($V_4$) correspond to Coulomb oscillations of QD$_4$ occupied by several electrons, yielding sufficient tunnel coupling to the source electrode to result in reflectometry contrast. 
(d) Peaks in $\VH$($V_3$) are Coulomb oscillations of QD$_3$ occupied by several electrons, yielding sufficient tunnel coupling to the source electrode (and sufficient capacitive coupling to G$_4$) to result in reflectometry contrast. 
For all panels $V_\mathrm{tg}$=6~V.
For compensated maps as in Fig.~\ref{fig3}f and Fig.~\ref{fig4}, voltage parameters $V_{1,2,3}^-$ ($V_{2,3,4}^+$) are defined such that they control the potential of QD$_1$, QD$_2$, and QD$_3$, respectively, without (with) affecting the potential of QD$_4$ (dashed arrows). 
}
	\label{fig2}
\end{figure}

Figure~\ref{fig2} shows stability diagrams for the longitudinal, diagonal, and transverse double quantum dots (DQD) indicated in its insets.
 Source and drain contacts are grounded for reflectometry measurements.
Discrete capacitive shifts of Coulomb peaks associated with one dot---serving as a charge sensor for the other dot---clearly reveal the threshold voltage for the first and second electron in all four dots.  

To achieve high-visibility Coulomb oscillations in $\VH$($V_4$), QD$_4$ is first configured in the multi-electron regime (9-12 electrons) \cite{Lundberg2020, CirianoTejel2021, Ibberson2021}.
Non-participating dots remain empty by applying 0~V to their gates. 
Multi-electron occupation increases the coupling strength of the $LC$ resonator to QD$_4$ and the tunnel rate between QD$_4$ and its reservoir, thereby facilitating dispersive sensing~\cite{Gonzalez-Zalba2015}. 
The large capacitive shift arising from the first and second electron on QD$_1$ (red dashed lines in Fig.~\ref{fig2}a) is qualitatively consistent with the relatively large dot-to-dot capacitance inferred from triple-dot measurements in Fig.~\ref{fig3} (see capacitance values in Appendix C). 

The reflectometry signal does not reveal single-electron charging of QD$_4$, likely due to its small tunnel rate relative to the reflectometry frequency (191~MHz)~\cite{Gonzalez-Zalba2015}. Presumably, other groups encountered the same limitation in similar devices and made no statements about single-electron occupation of their sensor dots~\cite{Chanrion2020, CirianoTejel2021}. 
In Figure~\ref{fig2}d, we therefore configure QD$_3$ in the multi-electron regime, yielding sufficient tunnel coupling to its reservoir and sufficient capacitive coupling to the reflectometry gate G$_\mathrm{4}$ to yield visible Coulomb oscillations in $\VH$($V_3$). In this way, discrete capacitive shifts of these Coulomb peaks indicate the first and second electron on QD$_4$ (black dashed lines). For this measurement, QD$_1$ and QD$_2$ were kept empty ($V_{1,2}$=-0.2~V). 

The threshold voltages observed in Fig.~\ref{fig2} for the first electron in each dot are not expected to be identical, due to different capacitive crosstalk between the highly-biased sensor gate (G$_4$ for panels a-c, G$_3$ for panel d) to the neighboring dots. After accounting for such capacitive effects (as done in the Supplementary Information of Ref.~\onlinecite{Ansaloni2020}), a spread of approximately 50~mV remains, comparable the observed spread among different cool downs of the same device. 
Recent simulations that go beyond the model in Appendix B suggest charged defects (such as dangling bonds) under or around the gates as a possible explanation~\cite{Martinez2021}. 

\subsection{Charge stability diagrams using compensated control voltages} 
\label{compensated}

The sensitivity of Coulomb oscillations of QD$_4$ (from now on referred to as sensor dot) to nearby charge rearrangements can be further utilized by applying compensated control voltages, i.e. linear combinations of native gate voltages $V_\mathrm{1,2,3,4}$ such that changes applied to compensated voltages do or do not change the chemical potential of the sensor dot \cite{Volk2019,Mills2019}. 
Such compensated control voltages are visualized in Fig.~\ref{fig2} as arrows.
Experimentally, they are implemented by calibrating capacitive matrix elements $\alpha_{4i}$ such that $V_\mathrm{4}$ compensates for electrostatic cross coupling between gates G$_\mathrm{1-3}$ and QD$_4$, i.e. by updating voltage $V_\mathrm{4}\equiv V^\mathrm{o}_\mathrm{4} +\sum_{\mathrm{i}=1}^{3} \alpha_{4i} (V_\mathrm{i}-V^\mathrm{o}_\mathrm{i}) $ whenever voltages $V_\mathrm{1-3}$ are changed relative to a chosen operating point $(V^\mathrm{o}_\mathrm{1},V^\mathrm{o}_\mathrm{2},V^\mathrm{o}_\mathrm{3})$. 
The choice of positive (negative) values for $\alpha_{4i}$ is indicated by adding a superscript + (-) to the respective control voltage, with $\alpha_{4i}$ listed in Appendix D. 
Using this compensation, and by setting the desired operating point of the sensor dot via $V_\mathrm{4}^\mathrm{o}$, the associated reflectometry signal  $V_\mathrm{H}$ becomes sensitive to charge rearrangements within the array. 

Positive compensation is useful for acquiring large stability diagrams of QD$_1$, QD$_2$, and QD$_3$, as it increases the density of Coulomb peaks associated with QD$_4$ and thereby facilitates the identification of charging events.
(This can be seen by comparing the density of sensor peaks in Fig.~\ref{fig4}a with that in Fig.~\ref{fig2}.)

Negative compensation, for accurate choices of $V^\mathrm{o}_i$ and $\alpha_{4i}$, has an intuitive physical interpretation: sitting on a Coulomb peak, as long as the (enhanced) reflectometry signal is unchanged, there are no charge rearrangements within the quadruple dot except for a continual exchange of electrons between QD$_4$ and its reservoir. 
This allows the study of charge state boundaries, demonstrated below for a triple-dot configuration, relevant, for instance, for the spatial permutation of isolated fermions~\cite{Ansaloni2020} or the implementation of exchange-only qubits~\cite{Medford2013}.

\begin{figure}[b]
\includegraphics[scale=1]{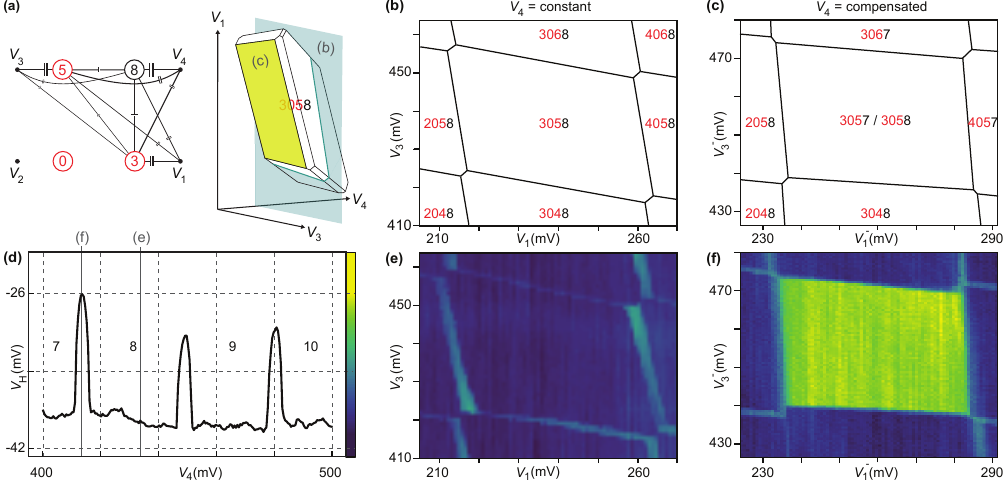}
\caption{\textbf{Uncompensated and compensated charge stability diagrams.} 
(a) Triple-dot circuit model of the quadruple-dot occupation 3058, and its calculated ground-state region. 
The boundary to 3057 (yellow) corresponds to charge transitions of the sensor dot (QD$_4$), whereas for fixed $V_4$ (cyan plane) QD$_4$ is generally in Coulomb blockade. The size of capacitor symbols represents their values used for simulations in panels b and c. 
(b) Cut through the stability diagram of (a) for $V_4=$~constant, revealing a hexagon reminiscent of double-dot behavior of QD$_1$ and QD$_3$.
(c) Cut through the stability diagram of (a) along the yellow plane, revealing a tetragonal region in which charge states 3057 and 3058 are degenerate. 
(d) Measurement of $V_\mathrm{H}$($V_4$), with other gate voltages fixed, showing an enhancement of the reflectometry signal at three sensor-dot transitions. 
(e) $V_\mathrm{H}$($V_1,V_3$) with $V_4$ fixed deep inside the 8-electron Coulomb valley as indicated in d. 
Apart from faint transitions associated with QD$_1$ and QD$_3$ (arising from capacitive coupling of G$_4$ to QD$_1$ and QD$_3$), no sensor-dot transitions are visible. 
(f) $V_\mathrm{H}$($V_1^-,V_3^-$), revealing a tetragonal region of enhanced reflectometry signal (QD$_4$ transitions). Here, the superscripts indicate that $V_4$ is negatively compensated when sweeping $V_1$ and $V_3$ (as illustrated in Fig.~\ref{fig4} a,c), thereby maintaining the 3057-3058 degeneracy of the sensor dot (as indicated in d) within the tetragonal region. 
$V_\mathrm{tg}$=12~V and $V_2$=0~V. 
Panels e and f use colorscale in d.    
Red and black numbers denote occupation of QD$_{1,2,3}$ and QD$_4$, respectively. 
  }	
	\label{fig3}
\end{figure}

Figure~\ref{fig3}a shows a capacitive circuit model for the charge occupation 3058, and its simulated ground-state region in gate-voltage space. 
Here, the $i$th digit indicates the number of electrons on the $i$th dot. 
In this model, the empty dot (QD$_2$) is ignored, as its physical presence can be absorbed into the capacitance values of the triple-dot circuit.  
The triple-dot constant interaction model is appropriate for sufficiently small tunnel couplings within the array~\cite{vanderWiel2002}. 
(For dispersive reflectometry, we like interdot tunnel couplings to be sufficiently large to show up in the reflectometry signal, hence we occupy QD$_1$ (QD$_3$) by 3 (5) electrons, rather then configuring them in the one-electron regimes.) 
Points representing the 3057-3058 ground-state degeneracy are shaded in yellow, indicating that a compensated scan of this region is expected to yield enhanced reflectometry signals arising from QD$_4$. 
In contrast, a native cut at fixed $V_4$ is shaded in blue, indicating that a low reflectometry background is expected due to all ground-states in this region having a fixed charge on QD$_4$. 
Using capacitance values inferred from measurements, simulations reveal that ground-state boundaries can appear hexagonal (as in the native plane of Fig.~\ref{fig3}b) or tetragonal (as in the negatively compensated plane of Fig.~\ref{fig3}c). 

To verify this observation experimentally, we set $V^\mathrm{o}_i$ and $\alpha_{4i}$ appropriate for configuring the device in the 3058 occupation and initially acquire uncompensated Coulomb oscillations associated with QD$_4$ (Fig.~\ref{fig3}d). Fixing $V_4$ inside the 8th Coulomb valley then yields uncompensated stability diagrams as in Fig.~\ref{fig3}e. As expected, no Coulomb peaks of QD$_4$ are visible (note colorscale in Fig.~\ref{fig3}d), but faint hexagonal features are clearly visible that correspond to smaller dispersive signals arising from charge transitions of QD$_1$ and QD$_3$~\footnote{It is for this illustrative reason that we show measurements for the 3058 charge state. For the 1018 state, the dispersive signals arising from charge transitions of QD$_1$ and QD$_3$ are too small to be visible}. 
In contrast, choosing an operating point on the 7th sensor peak yields negatively compensated charge stability diagrams as in Fig.~\ref{fig3}f, dominated by a tetragonal region with a $V_\mathrm{H}$ intensity consistent with the sensor peak indicated in Fig.~\ref{fig3}d. 

Well-known for capacitively coupled triple dots~\cite{Gaudreau2006,Hamo2016}, some charge state boundaries cannot be crossed by one-electron transitions alone and require two single-electron movements or two-electron cotunneling events, such as transitions 2058-3067 and 3048-4057 in Fig.~\ref{fig3}c. Surprisingly, these higher-order multi-electron dynamics clearly manifest themselves in the dispersive signal, at least for the high top-gate voltage used in Fig.~\ref{fig3}f. 

\begin{figure}[b]
	\includegraphics[scale=0.9]{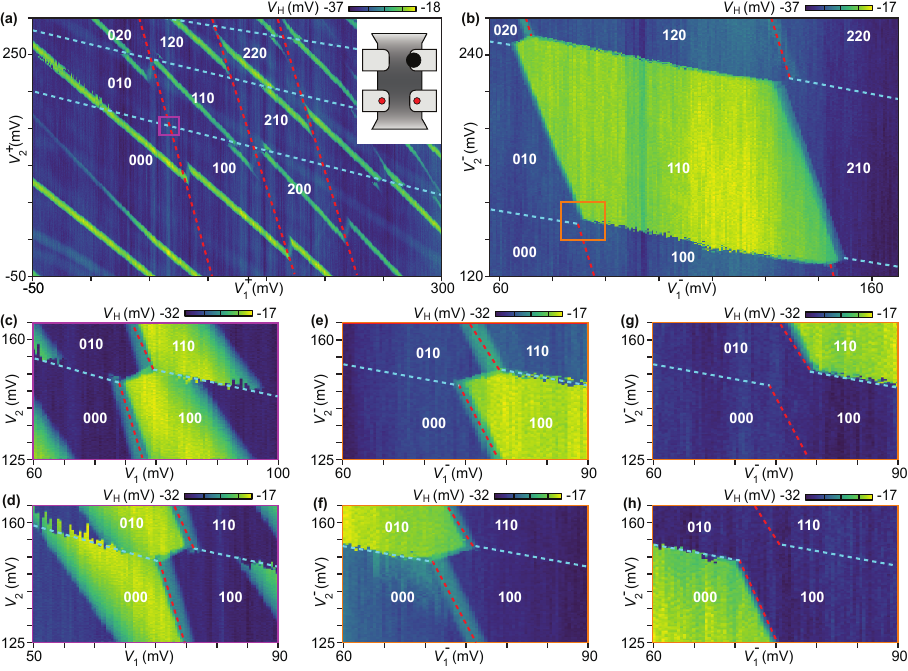}
	\caption{\textbf{Reflectometry with positive, negative, and without sensor-dot compensation.}
(a) Charge stability diagram of the transverse double dot (red dots, inset), measured with positively compensated sensor dot. 
Without positive compensation, the density of Coulomb peaks associated with the sensor dot would be significantly smaller and thereby make transitions of QD$_1$ and QD$_2$ less obvious (dashed lines). 
(b) Charge stability diagram of the same double dot, measured with negatively compensated sensor dot. 
The appearance of a hexagon (opposed to a tetragon as in Fig.~\ref{fig3}f) suggests that the transverse and diagonal double-dot configurations are capacitively qualitatively distinct.
(c)-(d) Charge transitions marked in (a), measured without compensation for two slightly different $V_4$ values. 
(e)-(h) Charge transitions marked in (b), measured with negative compensation for four different choices of the sensor operating point $V_\mathrm{4}^\mathrm{o}$. 
The drastically different charge-state contrasts evident in (c-h) suggest that the sensor-dot compensation should be chosen carefully depending on the application.
For all panels, QD$_3$ is empty by setting $V_{\mathrm{3}}$=0 and $V_{\mathrm{tg}}$=12~V. 
White numbers indicate occupation of QD$_{1,2,3}$. 
The occupation number of the sensor dot (QD$_4$) fluctuates in regions of high $V_\mathrm{H}$, and is not shown.
}
	\label{fig4}
\end{figure}

To show the practical differences between positive and negative compensation, we plot in Fig.~\ref{fig4}b a negatively compensated charge stability map of a transverse double dot (i.e. QD$_3$ deactivated) and in Fig.~\ref{fig4}a a positively compensated map containing the same charge states (in both cases, QD$_4$ is operated in the multi-electron regime to enhance $V_\mathrm{H}$). 
The high density of sensor peaks in Fig.~\ref{fig4}a makes it easy locate the degeneracy points associated with QD$_1$ and QD$_2$ (dashed lines), whereas the exact boundaries of individual charge states remain elusive. 
In contrast, Figure~\ref{fig4}b yields the exact shape of a particular charge-state boundary, from which device capacitances analogous to Fig.~\ref{fig3}a can be extracted. 

The observation of a hexagonal region in Figure~\ref{fig4}b, opposed to a tetragonal region as in Fig.~\ref{fig3}f, indicates that the two array configurations are represented by triple-dot circuits that are qualitatively different in terms of their effective capacitances. 

Compared to double dots with proximal charge sensors~\cite{Volk2019a}, which can also be viewed as triple dots, the relatively strong capacitive coupling between sensor dot and other dots in the 2$\times$2 array also makes \emph{uncompensated} charge stability diagrams qualitatively different, in the sense that only two or three charge states near a double-dot triple point can be distinguished. 
This is evident in Figure~\ref{fig4}c, where certain charge transitions like 000-010 cannot be distinguished despite the use of an intentionally power-broadened sensor peak. 
This sensitivity is useful when only one charge transition needs to be detected, in which case its visibility can be optimized or even reversed by adjusting the sensor operating point. 
For example, raising $V_\mathrm{4}$ by only 4~mV yields the opposite $V_\mathrm{H}$ contrast for the same 100-010 transition (Fig.~\ref{fig4}d). 

Negative compensation suffers from a similar ``strong coupling'' problem, making it difficult to distinguish multiple charge states within one charge stability map. 
Between Figures~\ref{fig4}e and \ref{fig4}h, only $V^\mathrm{o}_4$ was adjusted, resulting in an enhancement of $V_\mathrm{H}$ for four different charge states. 
		
Recently, a different method to mitigate the strong capacitive coupling between dots achieved charge sensing by rastering native gate voltages and plotting the Coulomb-peak position of the sensor dot (quantified as a change in sensor-dot voltage)~\cite{Chanrion2020}.

\subsection{Pulsed-gate charge-relaxation measurements}	
\label{pulsegate}

The ability to acquire charge stability diagrams facilitates the construction of gate-voltage trajectories that manipulate the charge configuration of the quadruple dot on nanosecond time scales. High-frequency reflectometry performed concurrently with gate-voltage pulsing allows monitoring of the charge dynamics on microsecond time scales (see Appendix E for an analysis of the bandwith and signal-to-noise ratio achieved in this work). This allows detailed charge-relaxation experiments, suitable to detect Pauli spin blockade and investigate spin physics in such devices~\cite{Lundberg2020, CirianoTejel2021,Oakes2022}. Pauli spin blockade between two quantum dots typically manifests itself as asymmetric charge relaxation rates in small magnetic fields; in the simplest case, the (20) spin-singlet state of a DQD can transition quickly into a (11) state, whereas (11) spin-triplet states only slowly relax to (20) (here, the Pauli exclusion principle requires a spin-non-conserving transition). In more complicated cases, such as in multi-electron regimes, additional considerations may become important~\cite{Lundberg2020, Higginbotham2014}. 

To illustrate the implementation of pulsed-gate experiments that reveal bidirectional interdot relaxation dynamics, we activate G$_2$ and G$_3$ as a DQD and continue to use QD$_4$ as a sensor dot. To increase tunnel rates, we set $V_\mathrm{tg}$=+30~V and keep QD$_1$ empty by setting $V_1$=-0.4~V. 

\begin{figure}
\includegraphics[scale=1]{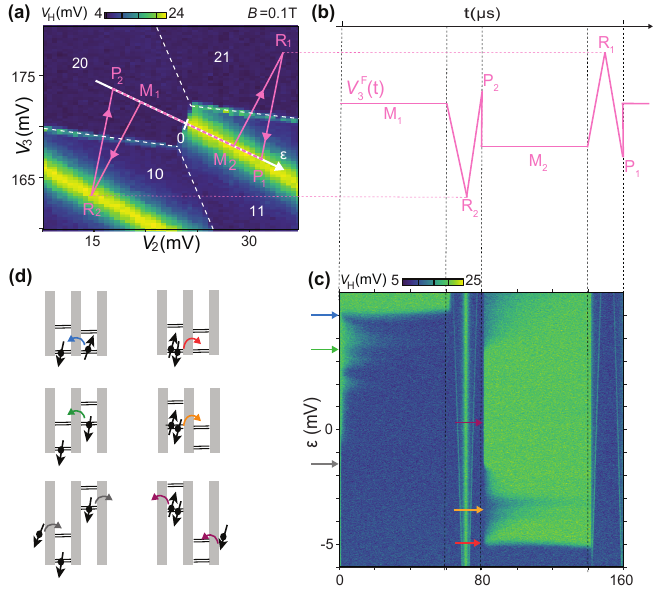}
\caption{\textbf{Pulsed-gate charge-relaxation measurements.} 
	(a) Uncompensated charge stability diagram for the DQD below G$_\mathrm{2}$ and G$_\mathrm{3}$, with an external magnetic field $B$=0.1 T applied parallel to the silicon channel. The $\varepsilon$-arrow represents a detuning range from -5 to +5 mV. A periodic 7-segment gate-voltage loop is shown by the pink trajectory.
	(b) Periodic $V_3^\mathrm{F}$(t) implementing the loop with a period of 160 $\mu s$. The concurrent $V_2^\mathrm{F}$(t) is constructed in a similar manner (not shown). 
	Ramps to the refresh (R$_1$ and R$_2$) and preparation points (P$_1$ and P$_2$) last 10~$\mu s$, while measurement segments at M$_1$ and M$_2$ last 60~$\mu s$.
	(c) Each row represents $V_\mathrm{H}(t)$, averaged over 500 gate-voltage loops, for a choice of DC operating values ($V_2, V_3$) specified by $\varepsilon$. The detuning dependence of (11)-to-(20) charge relaxation appears in the M$_1$ segment (0-60 $\mu s$), whereas reverse processes from (20) to (11) appear in the M$_2$ segment (80-140 $\mu s$).  
	(d) Interpretation of relaxation pathways for selected detunings in panel (c), based on energy-conserving tunneling from discrete Kramers doublets within one dot to another dot (see main text).  
}
\label{fig5}
\end{figure}

Figure~\ref{fig5}a shows the DQD charge stability diagram near the (20)-to-(11) interdot transition in the absence of gate-voltage pulses, using an uncompensated sensor dot. This diagram is used to define a detuning parameter, $\varepsilon$, as shown, with $\varepsilon$=0 defined at the ground-state degeneracy between (20) and (11). The voltage trajectory indicated by magenta arrows can be traced out in time by repeatedly applying suitable waveforms $V_2^\mathrm{F}(t)$ and $V_3^\mathrm{F}(t)$ (Fig.~\ref{fig5}b), while fixing the DC values $V_2$ and $V_3$ at $\varepsilon$=0. This works because the voltage trajectory is a closed loop (here with a period of 160~$\mu s$) with ramp times and ramp amplitudes chosen in such a symmetric manner that the DC blocks of the cryostat (i.e. capacitors $C_\mathrm{BT}$) do not introduce an effective time-averaged offset voltage on G$_2$ and G$_3$. Averaging $V_\mathrm{H}(t)$ over many gate-voltage loops then results in the row $\varepsilon$=0 of Fig.~\ref{fig5}c. Other rows in Fig.~\ref{fig5}c are obtained in a similar manner by stepping $V_2$ and $V_3$ along the $\varepsilon$-axis in ~\ref{fig5}a
~\footnote{$V_2=0.88 \varepsilon+V_2^0$, $V_3=-0.47 \varepsilon+V_3^0$, with $(V_2^0,V_3^0)$ centered at the inderdot transition.}. 
For each value of $\varepsilon$, $V_\mathrm{H}(t)$ is sampled at 100 MS/s and averaged over 500 loops. 

The bidirectional charge relaxation diagram in Fig.~\ref{fig5}c is of practical value. Between 60 and 80 $\mu s$, the gate-voltage trajectory ramps from the first measurement point in (20) (M$_1$) to the (10) configuration to refresh one electron (R$_2$), before preparing the system in the (20) configuration (P$_2$). The appearance of three sharp features within 60-80 $\mu s$ and their weak dependence on $\varepsilon$ is of diagnostic value, indicating, for example, that no charge switches or drift of the sensor dot occurred during these acquisitions. 
Physical insight into interdot relaxation mechanisms is provided by inspecting the $\varepsilon$-dependence of $V_\mathrm{H}(t)$ within the first measurement segment (0-60 $\mu s$, for (11)-to-(20) processes) and within the second measurement segment (80-140 $\mu s$, for reverse processes from (20) to (11)). For example, the ground-state to ground-state transition at M$_1$ (blue marker) appears equally fast as the ground-state to ground-state transition at M$_2$ (red marker), showing no sign of Pauli spin blockade. Similar fast ground-state to excited-state transitions appear at discrete values of $\varepsilon$ (see for example green and orange marker), indicating perhaps that the orbital level structure within the G$_2$ and G$_3$ dots are discrete and differ from each other. 
For detunings in between such relaxation ``hotspots'', relaxation is observed to be slower, likely due to inelastic decays involving  evanescent-wave Johnson noise or phonons~\cite{Chen2021}.  
The instantaneous relaxation near $\varepsilon$=0 (in the regions of the gray and maroon marker) likely arises not from interdot tunneling, but from relaxation via the source and drain reservoirs, as such processes are energetically allowed if $V_2$ and $V_3$ are chosen sufficiently close to $\varepsilon$=0. This provides information about the relative heights of tunnel barriers that define the DQD. 

We have acquired charge relaxation diagrams similar to Fig.~\ref{fig5}c for various in-plane magnetic fields, applied parallel to the silicon channel, ranging from $B$=0 T to 2~T, with various ramp rates and ramp amplitudes, without observing any clear evidence for spin or valley physics. 

Overall, the spin-valley and orbital physics of quantum dots in etch-defined split-gate FDSOI devices is less explored and poorly understood relative to gate-defined planar SiMOS and Si/SiGe devices. 
Generally, small valley splittings make it more difficult to observe spin blockade, although progress in detection methods have been reported for Si/SiGe devices~\cite{Jones2019,Borjans2021}. 
One complication is the low symmetry of the confinement potential in FDSOI nanowires, arising from a combination of structural and electrical confinements (leading to the formation of ``corner'' dots as shown in Fig.~\ref{fig1S}). This makes the prediction of valley splittings more difficult, especially in the presence of disorder due to charge traps and interface roughness~\cite{Martinez2021}, as the valley(-orbit) mixing depends on details of the wave functions at the interatomic length scale~\cite{Culcer2010, Corna2018}. 
Another complication one may need to consider is the effect of Coulomb correlations when going from a singly occupied to a doubly occupied dot (as in the (11)-to-(20) transition). In particular, Coulomb repulsion tends to split apart electrons in doubly occupied elongated dots such as corner dots. Recent simulations show that the formation of such ``Wigner molecules'' mixes different single-particle orbitals much more effectively than different valleys~\cite{AbadilloUriel2021}. This significantly reduces singlet-triplet splittings that (in the noninteracting picture) are dominated by an orbital excitation, but has weaker effects on singlet-triplet splittings that are dominated by a valley excitation, unless there is strong enough valley-orbit coupling~\cite{ercan2021, AbadilloUriel2021}. To look for such effects experimentally, one could use pulsed-gate spectroscopy of excited states to systematically compare one-electron and two-electron orbital excitations of the same dot~\cite{Pecker2013}. 
For these reasons, we expect that the application of our reflectometry and pulsed-gate techniques to different charge occupations and to more samples with different gate lengths and tunnel rates can provide more physical insights and help the development of spin-qubit functionalities. 

\subsection{Charge sensing without reservoirs}	
\label{molecule}

\begin{figure}[h]
	\includegraphics[scale=0.99]{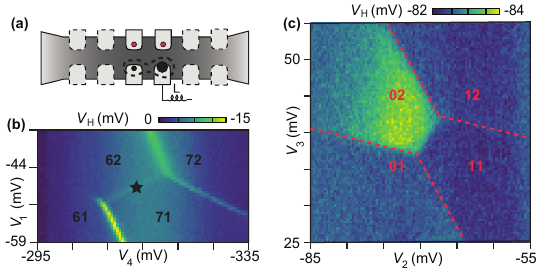}
	\caption{\textbf{Charge sensing without reservoirs.} 
	(a) Illustration of a 2xN quantum-dot device in which gate-reflectometry of an interdot quantum capacitance (black double-dot molecule) reveals the charge configuration of nearby spin qubits (red dots), thereby eliminating the need for sensor reservoirs. 
	(b) Uncompensated stability diagram of a double-dot molecule below G$_1$ and G$_4$ of our 2x2 device, with the double-dot occupation indicated by black numbers. Multi-electron occupation of QD$_4$ and $V_\mathrm{tg}$=30~V facilitate significant double-dot hybridization. The resulting interdot transition (black star) does not involve electron exchange with the reservoirs, but appears bright due to a dispersive shift of the reflectometry signal by the interdot quantum capacitance.  
	(c) Stability diagram of the two qubit dots below G$_2$ and G$_3$, obtained by fixing $V_1$ and $V_4$ at the interdot transition of the sensor double dot (black star in in (b)). The charge occupations of QD$_2$ and QD$_3$ (red numbers) in this regime where confirmed independently by using QD$_4$ as a conventional sensor dot. 
Acquisition of panel (a) and (b) used different settings for the reflectometry setup, resulting in an overall change of $V_\mathrm{H}$. 	
}
\label{fig6}
\end{figure}

When scaling from 2x2 devices to longer 2xN arrays, the source and drain reservoirs will eventually be too distant to support charge sensing within the bulk of the array. We address this challenge by demonstrating that charge sensing is possible without exchanging electrons with the leads. Our technique is based on creating a hybridized double dot within the array (Fig.~\ref{fig6}a),  
whose quantum capacitance is sensitive to nearby charges and can be detected as a dispersive shift in the reflectometry signal~\cite{Petersson2010}. We show this in our 2x2 device by activating two dots as a sensing DQD and the other two dots as qubit dots. 

To create the sensing DQD, the top gate is set to +30 V and QD$_4$ is populated by 6-7 electrons. In this regime, the interdot transition between QD$_4$ and QD$_1$ hybridizes charge states on both dots and gives rise to an enhanced reflectometry signal (black star in Fig.~\ref{fig6}b). Its sensitivity to nearby charges becomes evident by sweeping G$_2$ vs. G$_3$, as done in Fig.~\ref{fig6}c. The observed honeycomb pattern in $V_\mathrm{H}$ indicates that the sensor DQD not only senses changes to the total charge in the qubit array (note the strong contrast between (02) and (12), for example), but also inter-qubit charge transitions (such as (02) to (11)).  

Ultimately, future 2xN devices may benefit from reconfigurable dots, serving as qubit sites at some times and employed for readout or charge sensing at other times. Our gate-based DQD reflectometry technique may simplify such applications, as it does not require proximal reservoirs or dedicated sensor dots. 
	
\section{Conclusion} 
\label{conclusions}

This work demonstrates gate-based reflectometry measurements of various few-electron charge states in a 2$\times$2 quadruple dot implemented by 300-mm-wafer foundry fabrication. 
The strong mutual capacitances within the densely-packed (64-nm gate pitch) array of silicon quantum dots allows detection of single-electron tunneling in all four dots using only a single $LC$ resonator, wirebonded to one of the four side gates and monitored by radio-frequency reflectometry. 

Positive and negative compensation of the sensor dot potential yields convenient multi-dot stability diagrams with qualitatively distinct charge-state polytopes, as exemplified for a triple-dot configuration. 
Application of periodic, symmetric gate-voltage loops allow the acquisition of charge relaxation diagrams containing forward and reverse interdot tunneling within the same acquisition. This may help in the search for Pauli rectification and spin-valley selection rules, although in the present sample we have not found evidence for Pauli blockade. 
Finally, we demonstrate that the voltage-dependent hybridization between dots can be used to detect charge states of other dots, providing a route towards gate-based charge sensing in large 2xN arrays that does not require electron reservoirs or dedicated sensor dots.  


Further improvements in bandwidth, signal-to-noise ratio, and scalability may be possible by the use of Josephson parametric amplifiers~\cite{Stehlik2015,Schaal2020}, better impedance matching~\cite{Noiri2021a}, or integration with cryogenic control electronics~\cite{Clapera2015,Schaal2019}. 
Improved device geometries may harness individual control of tunnel barriers through advancements in three-dimensional very-large-scale integrated-circuit (3D VLSI) fabrication technologies~\cite{Vinet2018a}. 
Leveraging large gate capacitances and electrically-driven electron spin resonance~\cite{Corna2018} may then spark diverse applications for foundry-fabricated devices in circuit quantum electrodynamics~\cite{Ibberson2021}, quantum simulations, and spin-based quantum information processing.

\section{Acknowledgements}
We thank Silvano De Franceschi for the coordination of samples. This work received funding from EU grant agreements No. 951852, 688539, 676108, and 323841.
H.B. and F.A. contributed equally to this work.

\section{Appendices}

\subsection{Tunability of quadruple dot by top gate: modeling and experiment}
\label{tunability}

\begin{figure} [t!]
\includegraphics[scale=1.1]{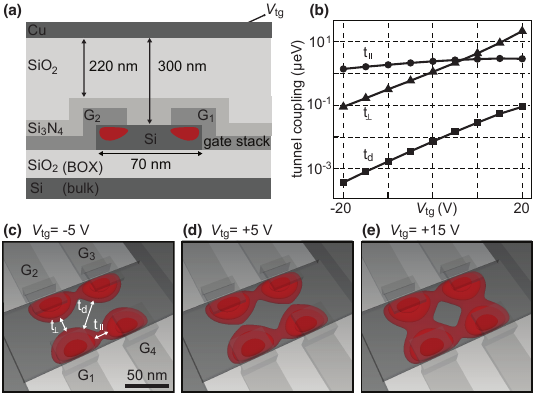}
\caption{\textbf{Dependence of simulated tunnel couplings on top-gate voltage.} 
(a) Schematic cross section including top gate (Cu) and location of quantum dots in the corners of the silicon nanowire (red). 
(b) Simulated parallel, transverse, and diagonal tunnel couplings (as indicated in c) as a function of $V_\mathrm{tg}$. 
(c)-(e) Contour plots of the computed ground-state electron wave functions for increasing $V_\mathrm{tg}$.
Dark, medium, and light red colors represent iso-surfaces at 0.001, 0.01, and 0.2 of the maximum electron density at $V_\mathrm{tg}$=0~V. }
\label{fig1S}
\end{figure}

We demonstrate theoretically and experimentally the overall tunability of tunneling rates between dots via the global top gate. 
Informed by the measured device (the schematic cross-section of the modeled device is shown in Figure~\ref{fig1S}a), our $\mathbf{k}\cdot\mathbf{p}$ model considers an accurately-sized nanowire together with its surroundings: source and drain reservoirs, gate electrodes, gate spacers and BOX substrate.
After self-consistently solving the potential in the device in the Thomas-Fermi approximation, the energies and wave functions of the tunnel-coupled single-electron quantum dots are computed with an anisotropic effective mass method (see \ref{detailskdotp}). 
Figure~\ref{fig1S}b shows the longitudinal $t_{||}$, transverse $t_{\perp}$, and diagonal $t_\mathrm{d}$ tunnel couplings as a function of top-gate voltage, with wavefunctions visualized in panels c-e. 

While $t_{\perp}$ and $t_\mathrm{{}d}$ strongly depend on $V_\mathrm{tg}$, $t_{||}$ shows a weaker dependence likely due to the short spacer $S_\mathrm{H}$ and the larger local screening arising from the geometry of adjacent wrap-around gate electrodes.  
Interestingly, $t_{\perp}$ is small at low top-gate voltages, becomes comparable to $t_{||}$ at $V_\mathrm{tg}\approx$5~V and is an order of magnitude larger at $V_\mathrm{tg}$=20~V, suggesting the use of global top gates to tune the ratio of transverse and longitudinal couplings ($t_{\perp}/t_{||}$). 

Independent simulations in Ref.~\onlinecite{Gilbert2020} for similar devices also found a significant effect of the top gate. 
Comparison between simulations and experimental data (tunneling times for different values of $V_\mathrm{tg}$ are reported in Ref.~\onlinecite{Ansaloni2020}) are, however, made difficult by the lack of knowledge of  the specific disorder (charged traps and interface roughness) under the spacers~\cite{Martinez2021}. The tunneling rates are, indeed, exponentially sensitive to fluctuations of the barrier height. The present simulations, which assume continuous distributions of charges at the interfaces, only account for the average effect of the traps on the potential in the barriers. A more quantitative modeling of the tunneling rates would require a detailed knowledge of the disorder in the particular device.


Our simulations also indicate large charging energies associated with each QD (17~meV) and large gate-coupling strengths (0.6~eV/V at $V_\mathrm{tg}$=0), which is attractive for high-temperature operation \cite{Yang2020, Petit2020}, dispersive gate sensing with high signal-to-noise ratios~\cite{Ansaloni2020, Chanrion2020}, and strong coupling to resonators or microwave cavities \cite{Ibberson2021}.
		
\begin{figure*}
\includegraphics[scale=1.05]{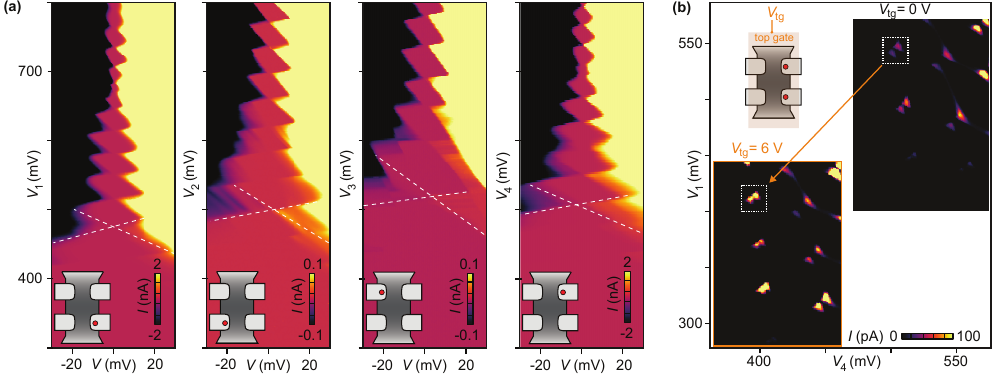}
\caption{\textbf{Transport measurements of single- and double-dot configurations.} 
(a) Coulomb diamonds for each quantum dot in the array, yielding charging energies of 15-20 meV in their few-electron regimes and 5-10 meV in their many-electron regimes. Gate lever arms are in the order of $\alpha\approx$0.5 eV/V. To allow such transport measurements, the gate in series with each QD gate is biased at high voltage. 
(b) The same set of bias triangles of a double-dot configuration measured for two different top-gate voltages, highlighting the overall increase of current for increasing top-gate voltage. 
In addition, all transport features shift to lower side-gate voltages when increasing $V_\mathrm{tg}$ (arrow), due to the capacitive cross coupling between top gate and quantum dots. $V_\mathrm{2,3}$=0~V and $V$=-3~mV for both $I$($V_1, V_4$) maps.  
	}
	\label{fig2S}
\end{figure*}

Experimentally, measurements of Coulomb diamonds as in Figure~\ref{fig2S}a reveal charging energies of 15-20~meV for the first few electrons, and gate strengths of 0.4-0.5~eV/V, for all four QDs, consistent with simulations. 
The tunability of tunnel rates by the top gate can be observed by a change of DC current, $I$, when the device is biased as a serial double dot:  
Figure~\ref{fig2S}b shows an increase in $I$ when the same bias triangles are measured at $V_\mathrm{tg}$=6~V instead of $V_\mathrm{tg}$=0~V. The apparent shift of these bias triangles by 100-125~mV towards lower values of $V_{1}$ and $V_{4}$ is consistent with a capacitive coupling of the top gate to dot potentials measured independently. 
Alternatively, characteristic tunnel times for different top-gate voltages can be measured in time domain using high-bandwidth reflectometry, as recently reported in Ref. \cite{Ansaloni2020}.

\subsection{Details of the $\mathbf{k}\cdot\mathbf{p}$ modeling}
\label{detailskdotp}

The device used for $\mathbf{k}\cdot\mathbf{p}$ modeling comprises a silicon channel ($W=70$~nm, $t_\mathrm{Si}=7$~nm, $L_{\mathrm{NW}}=165$~nm) with gate lengths $L_\mathrm{G}=32$~nm and gate spacings $S_\mathrm{V}=S_\mathrm{H}=32$~nm, consistent with the measured device. The simulated gate stack consists of 6 nm of SiO$_2$, 5 nm of TiN and 45 nm of poly-Si. 
The gate electrodes are capped by 25 nm of Si$_3$N$_4$ on each side, 
which mimic similar caps in the measured device that protect the channel during the ion implantation of source/drain dopants.
The whole device is encapsulated in SiO$_2$, with the 200-nm long top gate running 300 nm above the channel. 

To capture electrostatic screening by the reservoirs, 20-nm raised source and drain contacts have been added to both ends of the channel. They are highly $n$-doped ($N_\mathrm{d}=10^{20}$ cm$^{-3}$). 
Along the channel, the density of donors decreases by one order of magnitude every 4 nm from the outer edges of the source/drain spacers. 
Therefore, the regions underneath the gates and underneath the central spacers are practically undoped. 
The poly-Si gate is also $n$-doped ($N_\mathrm{d}=2\times 10^{19}$ cm$^{-3}$), while the silicon substrate below the 145-nm thick BOX layer is slightly $p$-doped ($N_\mathrm{a}=10^{15}$ cm$^{-3}$). We account for a 0.25~eV Schottky barrier at the interface between the poly-Si and TiN gates, inferred from the threshold voltage shifts measured at room temperature in similar devices with polysilicon-only gates. The dielectric constants of the materials are $\epsilon_{\rm{Si}}=11.7$, $\epsilon_{\rm{SiO}_{2}}=3.9$, and $\epsilon_{\rm{Si}_{3}\rm{N}_{4}}=7.5$. 
TiN is modeled as a perfect metal.

Charge traps in amorphous materials such as $\rm{Si}_{3}\rm{N}_{4}$ can shift the average potential and introduce disorder in the barriers. In order to capture this potential shift, we have modeled these traps as a continuous distribution of charges at the $\rm{SiO}_{2}/\rm{Si}_{3}\rm{N}_{4}$ interface. 
We observe that such majority carrier traps can significantly reduce the tunneling rates but have little impact on their tunability via the top gate. 
The results shown in Fig.~\ref{fig1S} assume a charge density of $-5\times 10^{11}$ $e\cdot$cm$^{-2}$ at the $\rm{SiO}_{2}/\rm{Si}_{3}\rm{N}_{4}$ interface.

The potential in the device is computed self-consistently within the Thomas-Fermi approximation. For numerical convenience, we assume a temperature $T=4.2$~K and account for incomplete ionization of the dopants at this temperature \cite{Altermatt2006}. 
The one-particle states in the ground-state Z valley are calculated with a finite differences implementation of the anisotropic effective mass approximation \cite{Venitucci2019}. 

We sweep the top-gate potential with the source, drain, and back-gate grounded. We apply the same voltage on all side gates G$_1$-G$_4$ such that the ground-state energy of the four-dot system remains resonant with the chemical potential of the source and drain. We then map the energies and wave functions of the four lowest-lying states onto the following effective Hamiltonian:
\begin{equation}
	\label{eq1}
	H = 
	\begin{bmatrix}
	E_\mathrm{Q1} & t_{\perp} & t_{d} & t_{||}\\
	t_{\perp} & E_\mathrm{Q2} & t_{||} & t_{d}\\
	t_{d} & t_{||} & E_\mathrm{Q3} & t_{\perp}\\
	t_{||} & t_{d} & t_{\perp} & E_\mathrm{Q4}
	\end{bmatrix}	
\end{equation}
where $E_{\mathrm{Q}i}$ are the energies of the isolated QDs, $t_{||}$ is the tunnel coupling between neighbouring QDs along the channel, $t_{\perp}$ is the tunnel coupling between opposite face-to-face QDs, and $t_\mathrm{d}$ is the tunnel coupling between diagonal QDs. With the same voltage on gates G$_1$-G$_4$, the system remains at a degeneracy point where $E_\mathrm{Q1} = E_\mathrm{Q2} = E_\mathrm{Q3} = E_\mathrm{Q4} = E_{0}$. 
The eigenenergies and parity (sign) of the wave functions in each dot are therefore:
\begin{subequations}
 \begin{align}
	\label{eq2}
	E_{1}&= E_{0} +  t_{||} + t_{\perp} +  t_{d};\,\psi_{1} = [+1,+1,+1,+1] \\
	E_{2}&= E_{0} +  t_{||} - t_{\perp} -  t_{d};\,\psi_{2} = [+1,-1,-1,+1] \\
	E_{3}&= E_{0} -  t_{||} + t_{\perp} -  t_{d};\,\psi_{3} = [+1,+1,-1,-1] \\
	E_{4}&= E_{0} -  t_{||} - t_{\perp} +  t_{d};\,\psi_{4} = [+1,-1,+1,-1]\, 
\end{align}
\end{subequations}
Once the calculated states have been unambiguously identified by their parities, $t_{||}$, $t_{\perp}$, and $t_\mathrm{d}$ can be fitted to their energies using the above equations. 

\subsection{Details of the constant-interaction capacitance model for the triple-dot configuration}

To simulate Figure~3b,c from the main text, we assume that the electrostatics of the triple-dot configuration can be described by a constant interaction model~\cite{vanderWiel2002} that is represented in Figure 3a as a circuit of 12 capacitors.
For sufficiently small tunnel couplings, this approximation is expected to be sufficient to capture the ground-state geometry (in gate-voltage space) of a particular charge configuration. 
Since QD$_\mathrm{2}$ was not activated in the experiment by setting $V_2$ to 0~V, we use in Figure 3a a capacitance circuit that only involves QD$_\mathrm{1}$, QD$_\mathrm{3}$, and QD$_\mathrm{4}$. (In reality, geometric capacitances associated with QD$_\mathrm{2}$ and G$_\mathrm{2}$ will contribute to some of these effective circuit capacitances.) 
To simulate the Coulomb valley of the 3058 configuration, we use the following capacitance matrix inferred from experimental stability diagrams:
	
\begin{equation}
	\label{eq3}
	C = 
	\begin{bmatrix}
	3.1 & - & 0.25 & 0.85\\
	- & - &- & -\\
	0.6 &- & 4.45 & 1.55\\
	0.75 & - & 0.4 & 5.5
	\end{bmatrix}	
\end{equation}
where diagonal elements $C_\mathrm{ii}$ correspond to the capacitive coupling between gate G$_\mathrm{i}$ and dot QD$_\mathrm{i}$, and off-diagonal elements $C_\mathrm{ij}$ correspond to the capacitive coupling between gate G$_\mathrm{j}$ and dot QD$_\mathrm{i}$.  
All capacitances are given in units of aF. 

In addition, the following dot-to-dot capacitances were used for the simulations in Figure 3b and 3c: 
QD$_\mathrm{1}$-to-QD$_\mathrm{3}$=0.25~aF, QD$_\mathrm{1}$-to-QD$_\mathrm{4}$=1.25~aF, QD$_\mathrm{3}$-to-QD$_\mathrm{4}$=0.75~aF. 
As indicated in Figure 3a, the smallest capacitance in the circuit is the ``diagonal'' capacitance between QD$_\mathrm{1}$ and gate G$_\mathrm{3}$ (0.25~aF), whereas the largest capacitance is the capacitance between QD$_\mathrm{4}$ and gate G$_\mathrm{4}$ (5.5~aF), consistent with its high occupation number. 

\subsection{Sensor operating points and compensation factors}
The gate-voltage compensations used to produce the figures are as follows.\\
Figure~\ref{fig3}f uses negative compensation: \\
$V_\mathrm{4}$ [V]$ = 0.4455 - 0.28 (V_\mathrm{1} - 0.26) - 0.115 (V_\mathrm{3} - 0.455)$\\
Figure~\ref{fig4}a uses positive compensation:\\
$V_\mathrm{4}$ [V]$ = 0.4 + 0.3 (V_\mathrm{1} + V_\mathrm{2})$\\
Figure~\ref{fig4}b uses negative compensation:\\
$V_\mathrm{4}$ [V]$ = 0.4455 - 0.28 (V_\mathrm{1} - 0.1) - 0.115 (V_\mathrm{2} - 0.18)$\\
Figure~\ref{fig4}e-h uses negative compensation:\\
$V_\mathrm{4}$ [V]$ = V_\mathrm{4}^\mathrm{o} - 0.28 (V_\mathrm{1} - 0.07) - 0.115 (V_\mathrm{2} - 0.15)$ \\
with slightly different values $V_\mathrm{4}^\mathrm{o}$ for each panel.\\

\subsection{Signal-to-noise ratio of charge sensing}
\begin{figure*}[h]
\includegraphics[scale=0.99]{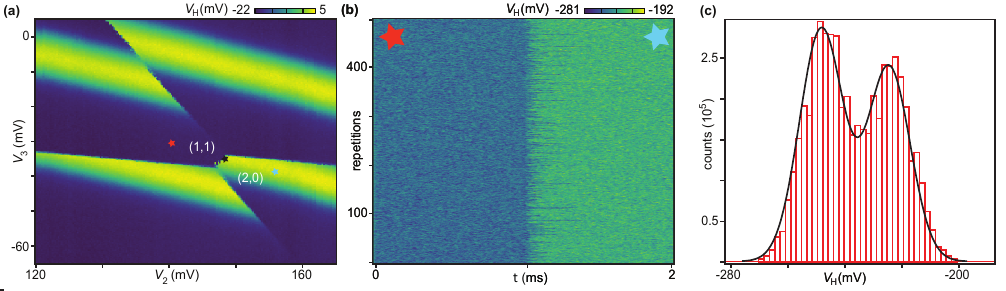}
\caption{\textbf{Signal-to-noise ratio of charge sensing.} 
(a) Stability diagram for the QD$_\mathrm{2}$-QD$_\mathrm{3}$ double dot at $V_\mathrm{tg}$ = 30~V. Repeatedly pulsing between (1,1) and (2,0) detuning points (red and blue marker), with a hold time of 1~ms at each point, results in a series of single-shot traces $V_\mathrm{H}(t)$ as shown in panel (b). 
(b) 500 single-shot traces measured by passing $V_\mathrm{H}(t)$ through a 30~kHz low-pass filter. 
(c) Histogram of all pixels in panel (b), fitted by two Gaussians, indicating a signal-to-noise ratio of 2.3.
 }
\label{fig3S}
\end{figure*}

In Figure~\ref{fig3S}a, a stability diagram around the (1,1)-(2,0) transition of the double dot under G$_2$ and G$_3$ is shown, for a top-gate voltage of +30 V and with QD$_1$ kept empty. 
Fixing the DC gate voltages $V_2$ and $V_3$ at the interdot transition (marked with a black star), the application of a square detuning pulse (50/50 duty cycle with a 2-ms period and an amplitude marked by a red and blue star) to $V_2^\mathrm{F}(t)$ and $V_3^\mathrm{F}(t)$ results in the data shown in Suppl. Fig.~\ref{fig3S}b. Each row represents a single-shot trace associated with one pulse period. For these measurements, the effective bandwidth of 30 kHz is set by passing the demodulated reflectometry signal $V_\mathrm{H}(t)$ through an analog low-pass filter (SRS model SR560) prior to digitization (AlazarTech ATS9360). Between acquiring panel (a) and (b), the gain of the SR560 and other settings in the reflectometry setup were adjusted, which explains the different ranges of $V_\mathrm{H}$. 
By converting panel (b) to a histogram and fitting it with the sum of two Gaussian functions as shown in Suppl. Fig.~\ref{fig3S}c, we extract the signal-to-noise ratio (SNR) as the offset $\Delta V_\mathrm{H}$ of the two Gaussians normalized by $\sqrt{\sigma_{1}^2 + \sigma_{2}^2}$, where $\sigma_1$ and $\sigma_2$ are standard deviations of the Gaussian functions. 
In this example, for a detection bandwidth of 30 kHz, the SNR is 2.3. 

\end{document}